\documentclass[referee,usenatbib]{mn2e}
\usepackage{graphicx,subfigure}
\usepackage[normalem]{ulem}
\usepackage{bm}
\usepackage{longtable}
\usepackage{amsmath,amssymb,graphicx}
\usepackage{rotating}
\usepackage{draftcopy}
\usepackage{color}
\newcommand\nat{Nature}%
\newcommand\apj{ApJ}%
\newcommand\aap{A\&A}%
\newcommand\apjl{ApJL}%
\newcommand\mnras{MNRAS}%
\newcommand\aj{AJ}%
\newcommand\physrep{PhR}%
\newcommand\thesis{Ph. D. Thesis}%
\newcommand{\Lb}{\left(}
\newcommand{\Rb}{\right)}

\draftcopyName{Draft/Version4}{80}

\title[Multi-wavelength modelling of GRB~050525A afterglow]
{Comprehensive multi-wavelength modelling of the afterglow of GRB~050525A}
\author[L. Resmi et al.]
{L. Resmi\thanks{e-mail :resmi@tifr.res.in}$^{1,2}$, K. Misra$^{3,4}$, 
G. J\'{o}hannesson$^{5}$, A. J. Castro-Tirado$^{6}$,   J. Gorosabel$^{6}$,
\newauthor
M. Jel\'{i}nek$^{6}$,  D. Bhattacharya$^{4}$,  P. Kub\'{a}nek$^{6}$, G. C. Anupama$^{7}$,   A. Sota$^{6}$,   
\newauthor
D. K. Sahu$^{7}$, A. de Ugarte Postigo$^{8}$,  S. B. Pandey$^{9}$, R. S\'{a}nchez-Ram\'{\i}rez$^{6}$,
\newauthor
 M. Bremer$^{10}$,  R. Sagar$^{9}$ 
\\
 1. Tata Institute of Fundamental Research, Mumbai 400 005, India \\
 2. Institut d'Astrophysique de Paris, Paris 75014, France \\
 3. Space Telescope Science Institute, Baltimore, Maryland 21218, USA \\
 4. Inter University Center for Astronomy and Astrophysics, Post Bag 4, Ganeshkhind, Pune 411 007, India \\
 5. Science Institute, University of Iceland, Dunhaga~3, IS--107 Reykjavik, Iceland \\
 6. Instituto de Astrof\'{i}sica de Andaluc\'{i}a (IAA-CSIC), PO Box 03004, 18080 Granada, Spain \\
 7. Indian Institute of Astrophysics, Koramangala, Bangalore 560 034, India\\
 8. Dark Cosmology Center,  Niels Bohr Institute,  Juliane Maries Vej 30,  Copenhagen, 2100, Denmark\\
 9. Aryabhatta Research Institute of observational sciencES, Manora Peak, Nainital 263 129, India \\
 10. Institute de Radioastronomie Millim\'etrique (IRAM), 300 rue de la Piscine, 38406 Saint Martin  H\`eres, France\\
}

\begin{document}


\date{Accepted.....; Received .....}

\pagerange{\pageref{firstpage}--\pageref{lastpage}} \pubyear{}

\maketitle

\label{firstpage}

\begin{abstract}
{The {\it Swift} era has posed a challenge to the standard blast-wave model of Gamma Ray Burst (GRB) afterglows. 
The key observational features expected within the model are rarely observed, such as the achromatic  steepening (`jet-break') of the light curves. The observed afterglow light curves showcase additional complex features requiring modifications within the standard model . 
Here we present optical/{\it NIR} observations,  millimeter upper limits and comprehensive broadband modelling of the afterglow of the bright GRB 0505025A,  detected by {\it Swift}.
This afterglow cannot be explained by the simplistic form of the standard blast-wave model. We attempt modelling the multi-wavelength light curves using (i) a forward-reverse shock model, (ii) a two-component outflow model and (iii) blast-wave model with a wind termination shock. The forward-reverse shock model cannot explain the evolution of the afterglow. The two component model is able to explain the average behaviour of the afterglow very well  but cannot reproduce the fluctuations in the early X-ray light curve. The wind termination shock model reproduces the early light curves well but deviates from the global behaviour of the late-time afterglow.
}
\end{abstract}



\section{INTRODUCTION}
\label{intro}

Gamma Ray Bursts (GRBs) are extremely energetic cosmic explosions which outshine 
the entire $\gamma$-ray sky for a few seconds. The launch of {\it Swift} (Gehrels
et al. 2004), a dedicated satellite to detect GRBs and rapidly follow-up their afterglow emission, has revolutionized the study of the most energetic cosmic explosions in the Universe.

In the standard blast-wave model for GRB afterglows (\citealt{1992MNRAS.258P..41R,1993ApJ...418L...5P}; also see \citealt{1999PhR...314..575P} for a review), a relativistic shock decelerates through uniform circumburst medium, heats up the matter, accelerates particles and enhances the magnetic field downstream. Synchrotron radiation from the shocked particles is observed as the afterglow. The snap-shot synchrotron spectrum can be characterized by four spectral parameters (apart from the electron index $p$): the injection frequency $\nu_m$, the cooling frequency $\nu_c$, the self-absorption frequency $\nu_a$, and the peak flux $f_m$. The spectral parameters can be mapped to four physical parameters: the isotropic equivalent energy $E_{\rm{iso}}$, the ambient medium density (parameterized as number density $n_0$ for a constant density medium and as $A_\star$ for a wind driven medium for which  $\rho(r) = 5.5 \times 10^{11} \frac{A_{\star}}{\rm{g \, cm}^{-1} } \Lb \frac{r}{\rm {cm} } \Rb^{-2}$ as in \citealt{1999ApJ...520L..29C}), and the fractional energy content in non-thermal electrons and magnetic field ($\epsilon_e$ and $\epsilon_B$ respectively). Jet-break, a simultaneous steepening seen in the multi-frequency light curves considered as a signature of the collimated outflow from the burst \citep{1999ApJ...525..737R}, if observed (at time $t_j$ since burst), gives a handle on the initial collimation angle ($\theta_j$) of the explosion, and hence to the total kinetic energy involved ($E_{\rm{tot}}$).

The model was largely successful in explaining the pre-{\it Swift} observations of GRB afterglows. 
However, {\it{Swift}} with its ability to locate the afterglow within minutes of the burst and follow it up in {\it UV}, optical and X-ray bands has revealed complexity in the early afterglow emission that is not predicted by the model. The X-ray light curves in the {\it Swift} era have been rather dramatic, with steep decays, plateaus and flares, not witnessed earlier \citep{2006ApJ...642..389N,2007ApJ...671.1903C}. 
Moreover, in several afterglows, the flux evolution did not follow the predicted spectral-temporal relations \citep{2008ApJ...675..528L}. This has led to the conclusion that afterglow light curves differ drastically from burst to burst , owing to various physical processes shaping the flux evolution in various bands \citep{2006ApJ...642..354Z}. Another open issue in the {\it Swift} era is the absence of a jet-break. These complications often make it a demanding task to extract the physics of the burst and its surroundings from afterglow data.

The bright low redshift ($z = 0.606$, \citealt{2005GCN..3483....1F}) Gamma Ray Burst 050525A was detected by the {\it Swift}-BAT on 2005 May 25 at 00:02:53 UT \citep{2005GCN..3466....1B}. We refer to the burst trigger time as $t_0$. An isotropic equivalent $\gamma$-ray energy of $2.3 \times 10^{52}$~erg is inferred for the observed BAT fluence at a distance of $3.57$~Gpc (assuming $\Omega_m = 0.3, \Omega_\Lambda = 0.7$ and $H_0 = 70$~km s$^{-1}$ Mpc$^{-1}$). The UVOT {\it V}-band observations started at $T=t-t_0\sim$65~s and XRT observations began 
$T\sim 75$~s leading to well sampled early afterglow light curves. The proximity of the burst and the brightness of the afterglow made it a very good target for multi-wavelength observations. Ground based optical observations including robotic telescopes \citep{2005A&A...439L..35K,2006ApJ...642L.103D}, radio observations in multiple frequencies by the Very Large Array \citep{2005GCN..3495....1C} and Spitzer observations at $\sim 2$~days in multiple {\it IR}-bands \citep{2008ApJ...681.1116H} have been reported in the literature.   However, a detailed modelling involving the full evolution of the relativistic shock to infer the physical parameters has not been attempted for this burst.

In this paper we present a new set of {\it VRIJH} observations using eight different optical telescopes and millimeter upper limits from IRAM Plateau de Bure Interferometer (PdBI).    We supplement our data with 
{\it Swift}-UVOT and XRT data reported by \cite{2006ApJ...637..901B}, the optical data reported by  \cite{2005A&A...439L..35K} and \cite{2006ApJ...642L.103D} and the radio data from the VLA afterglow repository\footnote{http://www.aoc.nrao.edu/$\sim$dfrail/allgrb\_table.shtml} to 
study the broad band evolution of the afterglow.  We model the afterglow using three different extensions of the standard blast-wave model e.g.  the forward-reverse shock model
\citep{1999MNRAS.306L..39M},  the two-component model \citep{2003Natur.426..154B} and a model including a wind termination shock
\citep{2006ApJ...643.1036P}. Section \ref{observations} gives a description of the data acquired from different telescopes and the analysis techniques. Multi-wavelength modelling under various premises are described in Section \ref{modelling}. Section \ref{conclusion} provides a summary of the multi-wavelength modelling results.

\section{Data acquisition and Reduction}
\label{observations}
\subsection{Millimeter wave observations}

The IRAM Plateau de Bure Interferometer \citep{1992A&A...262..624G}  (PdBI, France) observed the position of GRB 050525A simultaneously at 92.682~GHz and 214.712~GHz about one day after the burst (see Table \ref{table1}).  The fields of view of the 15-m antennas at these frequencies are respectively $54{\farcs}3$ and $23{\farcs}5$, and the synthesized beams in the compact 5-antenna configuration were $7{\farcs}05 \times 5{\farcs}58$ at $PA=74^\circ$ and $3{\farcs}10 \times 2{\farcs}36$ at $PA=78^\circ$ where the position angle $PA$ of the major beam axis is defined from North over East.
The data reduction was done with the GILDAS\footnote{http://www.iram.fr/IRAMFR/GILDAS}
software. We performed point source fits in the UV plane; for a fit fixed to the
phase center position, a signal of at least $3 \sigma$ is needed to claim a
detection, whereas a point source elsewhere in the field of view
must have at least $5 \sigma$. We did not detect a millimeter
counterpart, neither on the phase center coordinates nor elsewhere in
the field of view.  In the following, we take the
$3\sigma$ levels as upper limits, i.e. 1.02~mJy for 92.682~GHz and
3.66~mJy for 214.712~GHz (Table \ref{table1}).

\subsection{Optical and NIR observations of the afterglow}
\label{optical_data}
The afterglow of GRB~050525A was observed using 
different optical facilities in the broad bands $VRIJH$
during 2005 May 25 to July 01. We used 
the 0.2m BOOTES1-B telescope located at the INTA-CEDEA station at El Arenosillo, Huelva, Spain;
the 1.04m Sampurnanand Telescope (ST) in India;
the 1.2m semi-robotic Mercator telescope at La Palma;
the 1.5m telescope at the Observatorio de Sierra Nevada (OSN);
the 2.01m Himalayan Chandra Telescope (HCT) in India;
the 2.5m Issac Newton Telescope (INT)
 and
the 2.2m and 3.5m telescopes at the Calar Alto Observatory (CAHA) for the
observations (Table \ref{table2}). 
Several bias and twilight flat frames were acquired from different telescopes for pre-processing the CCD images. 
The pre-processing was done in a standard fashion 
including bias subtraction, flat fielding and cosmic ray removal in 
all object frames. 
Standard data reduction software ({\it IRAF}\footnote{{\it IRAF} stands for Image Reduction 
and Analysis Facility distributed by the National Optical Astronomy Observatories which is 
operated by the Association of Universities for research in Astronomy, Inc. under co-operative 
agreement with the National Science Foundation}, {\it MIDAS}\footnote{ {\it MIDAS} stands for Munich Image and Data Analysis System designed and developed by the European Southern Observatory {\it ESO} in Munich, Germany}, and {\it DAOPHOT}\footnote{ {\it DAOPHOT} stands for 
Dominion Astrophysical Observatory Photometry} \citep{1987PASP...99..191S}) 
were used for photometric analysis. The instrumental magnitudes of the optical afterglow were differentially calibrated using nearby secondary stars 
from the list of \cite{2005GCN..3855....1H}.
A full compilation of {\it{VRIJH}} magnitudes of the afterglow 
is presented in Table \ref{table2}.  

\begin{figure}
\begin{center}
\includegraphics[scale=0.5]{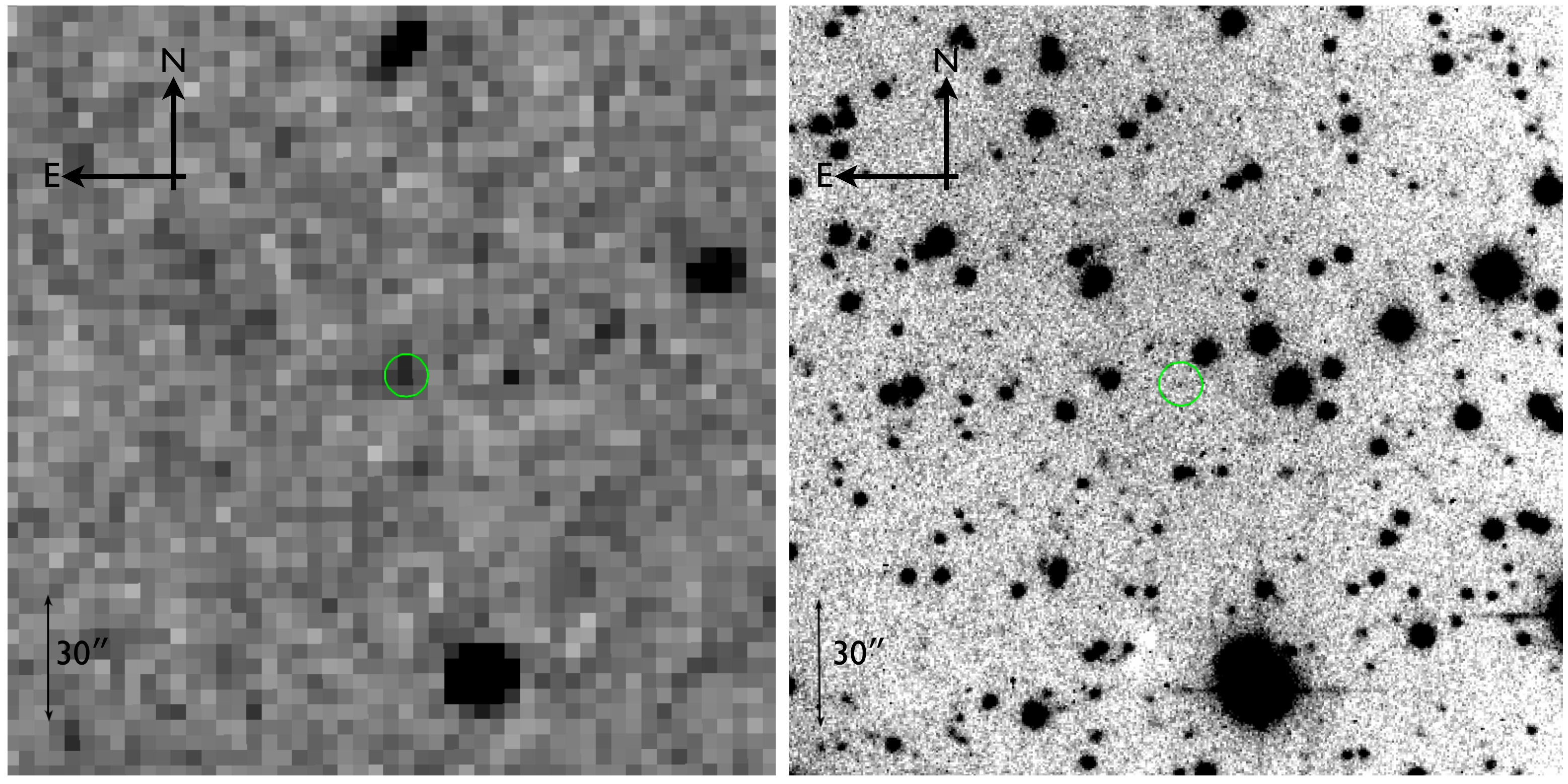}
\end{center}
\caption{The GRB~050525A field. The left image (at $T\sim$700 ~s after the burst,  {\it V}-band) was taken with the 0.2m telescope at the BOOTES1-B  telescope located at the INTA-CEDEA station at El Arenosillo, Huelva, Spain. The right one, shown for comparison purposes, was taken  at $T\sim$2~days after the burst with the 1.5m telescope at IAA-CSIC Observatorio de Sierra Nevada (OSN) in the {\it I}-band.  The green circle is 5$\farcs$0 in radius and indicates the location of the afterglow in the two images.}
\label{figure1}
\end{figure}

The earliest observations were taken with the BOOTES1-B telescope at $T \sim 0.0044$~days. We detect the afterglow in a co-added early time BOOTES1-B image.  
Figure \ref{figure1} shows the optical afterglow in the early time BOOTES1-B image and in an image taken $\sim 2$ days after the burst with the OSN telescope.   Late-time observations of the afterglow on 2005 June 30 and 2005 July 01 with the 2.2-m CAHA and the 2.5-m INT respectively resulted in a non-detection and an upper limit.

\begin{figure}
\begin{center}
\includegraphics[scale=0.9]{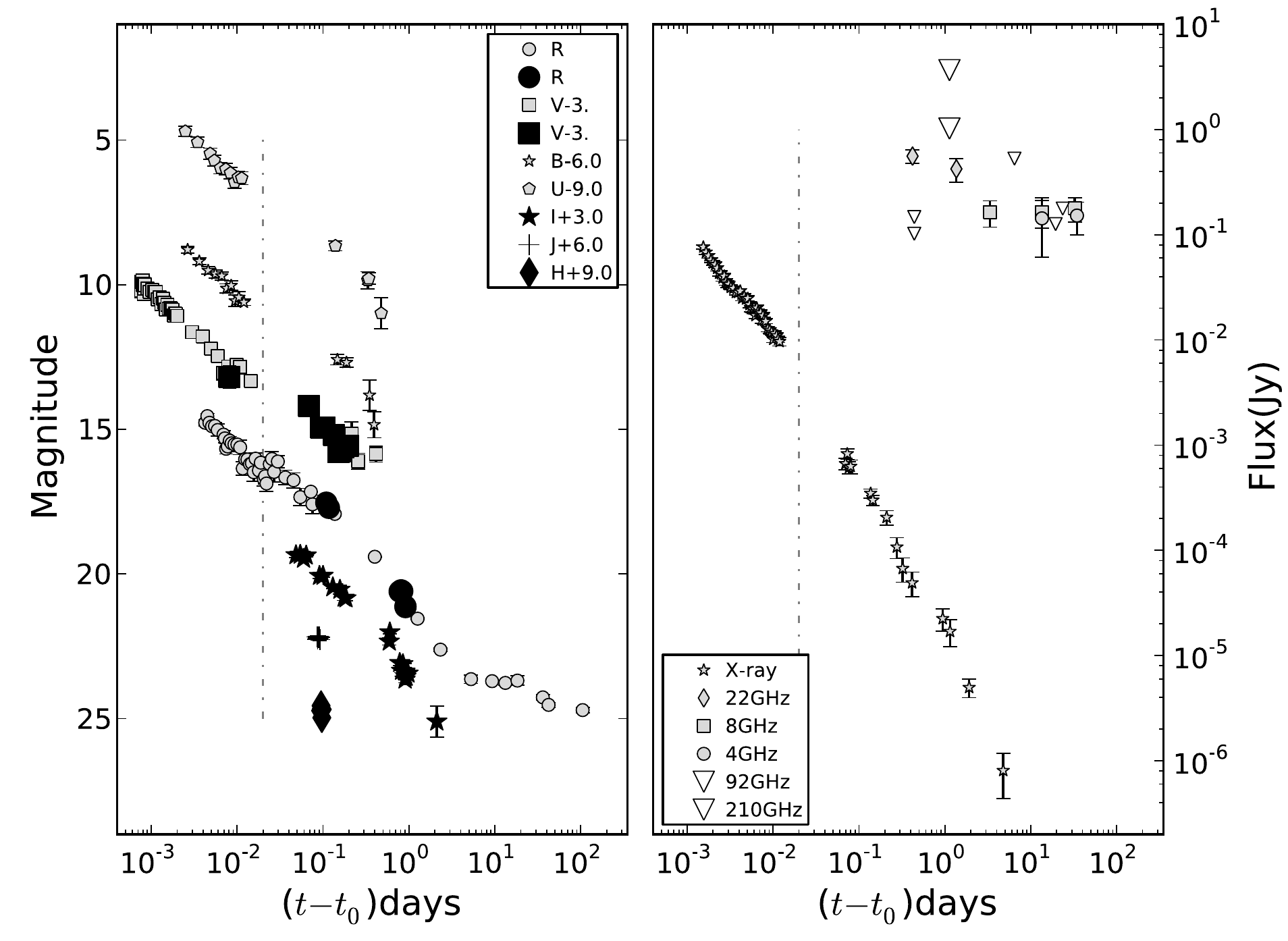}
\caption{{\it UBVRIJH}, X-ray,  radio and millimeter light curves of the afterglow of GRB~050525A.
The dark and light coloured symbols correspond to the new data presented in this paper and the literature data respectively. The upper limits, including the millimeter ones presented in this paper, are represented as open triangles. The dash-dotted vertical line corresponds to $T \sim 0.02$~days (see text). The optical light curves are shifted arbitrarily for clarity.}
\label{figure3}
\end{center}
\end{figure}

In Figure \ref{figure3} we show the optical/NIR and mm data presented in this paper along with the published data in X-ray, optical and radio frequencies. 
We present an {\it I}-band light curve for this burst for the first time, which extends from $T \sim 0.05$ to $0.6$~days and our data fills the gap in {\it V}-band light curve between $0.05$ -- $0.2$~days ($T\sim5800$ -- $17,000$~s).  Our infrared observations ({\it J-} \& {\it H-}bands) are clustered around $0.09$~days, but provide spectral information within the optical band that constrains model parameters. 


\subsection{Host galaxy observations}
\begin{figure}
\begin{center}
\includegraphics[scale=0.4]{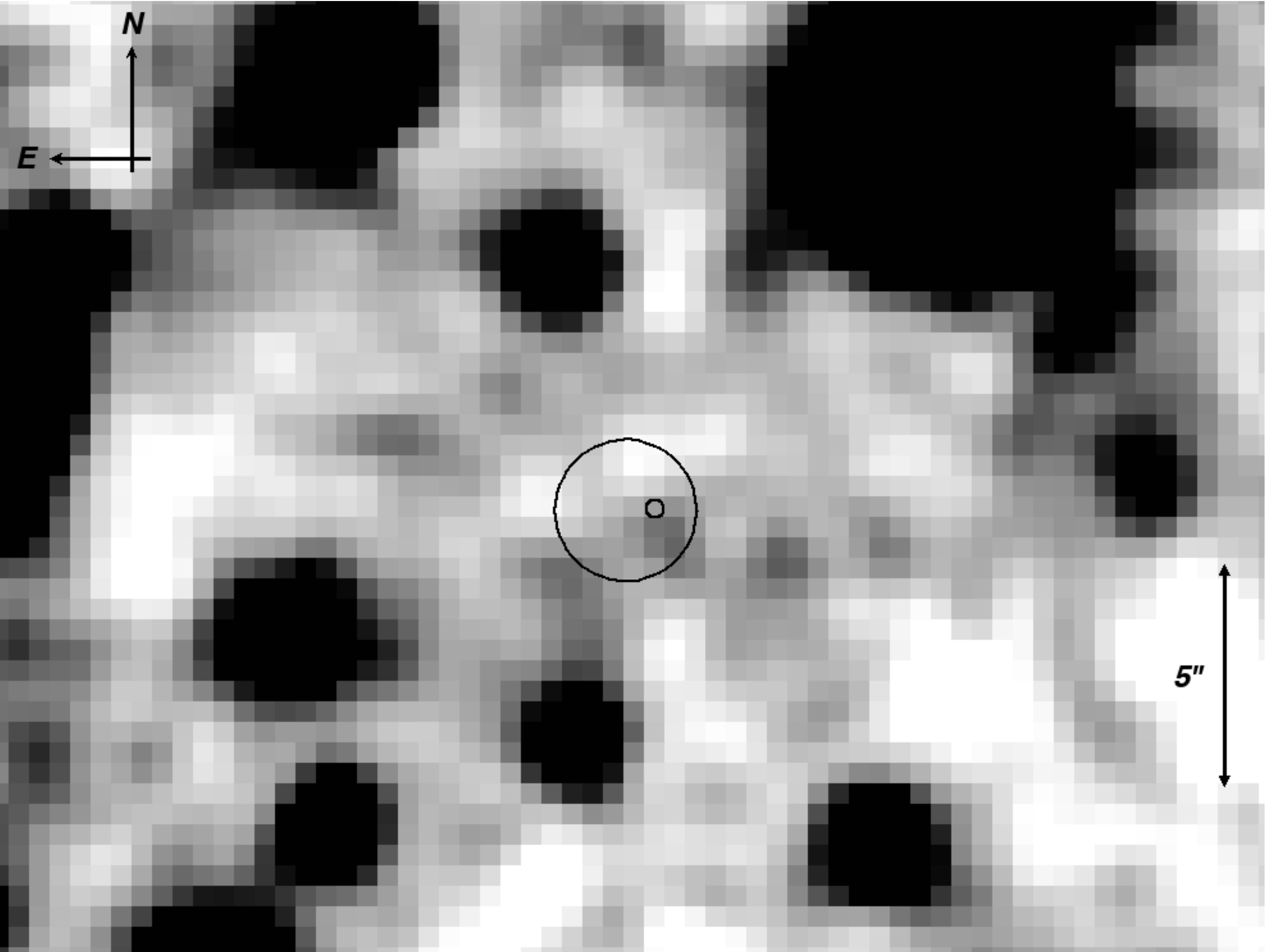}
\end{center}
\caption{A late-time $J$ band image of the GRB~050525A field with the 3.5m CAHA telescope.
We detect four faint sources close to the {\it Swift} XRT error circle. The source closest to the enhanced XRT position (90\% error circle) and UVOT position (1-$\sigma$ error circle) reported in \citet{2006ApJ...637..901B} is indicated in the figure and is assumed to be the GRB host galaxy.  The photometry of the host galaxy is presented in Table \ref{table2}.}
\label{figure2}
\end{figure}

We obtained deep imaging of the GRB~050525A field in the {\it gri} bands between 2011 May 26--28 with the 2.2m CAHA telescope.  The data reduction was carried out in a manner similar to the one described in Section \ref{optical_data}.
A very faint source is detected in the $i$ band at the location of the afterglow but only upper limits can be imposed in the $g$ and $r$ bands.  The log of the observations is given in Table \ref{table2}.
We attempted further deep imaging in the $J$ band on 2011 July 18--19 for a total exposure time of 2.33 hours spread over two consecutive nights with the 3.5m CAHA telescope equipped with the near-IR camera Omega$_{2000}$.  The photometric calibration is carried out by observing the UKIRT faint standards FS28 and FS30. 
We detect four faint sources close to the {\it Swift} XRT error circle in the $J$-band image.
The source closest to the enhanced XRT 
position\footnote{http://www.swift.ac.uk/xrt\_positions/} and UVOT position reported in \cite{2006ApJ...637..901B} is indicated in 
Figure \ref{figure2} and is assumed to be the host galaxy of GRB 050525A.  
The host galaxy photometry is presented in Table \ref{table2}.

\section{Multi-Wavelength Modelling Results}
\label{modelling}
GRB~050525A has a relatively well sampled afterglow light curve (see Figure \ref{figure3}). The {\it Swift} observations end at $T\sim1$~day while ground based {\it R}-band observations continue up to $\sim20$~days. The early X-ray light curve starts off with a  powerlaw decay but a minor flare is observed at $T\sim0.0035$~days \citep{2006ApJ...637..901B}. This flare is not seen in the optical light curves that undergo a smooth powerlaw decay. However, around $T\sim0.02$~days, a re-brightening is observed in the {\it R}-band light curve. 
There is considerable scatter in the {\it R}-band light curve at this epoch and unfortunately no other frequencies cover this epoch. The afterglow evolution after this period appears to be within the expectations of the standard blast-wave model with a near simultaneous break and similar post break indices in the optical and X-ray light curves \citep{2006ApJ...637..901B}. However, the light curve before $0.02$~days significantly deviates from the extrapolation of the powerlaw decay that is present afterwards. 
There is a change in the $R$-band light curve normalization after this epoch, or in other words, an increase in optical to X-ray spectral index with no color evolution within the X-ray or optical bands. This suggests that the value of $\nu_c$ decreases rapidly in this period. A single component, like a forward shock emission, alone cannot explain the complete evolution of this afterglow. Either there are multiple emission components or there is a density discontinuity in the ambient medium. 

We perform simultaneous multi-wavelength fitting of the afterglow data to obtain the underlying physical parameters. Parameterized extinction laws for Milky Way \citep{1989ApJ...345..245C}, LMC, SMC \citep{1992ApJ...395..130P} and starburst galaxy \citep{1997AJ....113..162C} are used for modelling extinction due to dust column in the host galaxy.
A chi-square minimization procedure is used to derive the best fit parameters.  
For fitting we use all data except 
the late-time ($T >5$~days) data in $R$-band that is possibly contaminated by a supernova (SN) component (see section \ref{sn}).

For the forward-reverse shock model and the two-component model, our code calculates the synchrotron flux $f_\nu(t)$ at an observed frequency $\nu$ at a given time $t$ as a function of the four spectral parameters which evolve following the respective shock dynamics \citep{1999ApJ...523..177W, 2008MNRAS.388..144R}. We assume both wind driven and constant density ambient medium while searching for the best fit. The free parameters of forward-reverse shock model and two-component model are the input values of $\nu_m$, $\nu_c$, $\nu_a$ and $f_m$ at a given epoch, the jet-break time $t_j$, electron index $p$ and the dust extinction $E(B-V)$ in the host galaxy. If the electron distribution is a hard powerlaw, the synchrotron frequency $\nu_i$ corresponding to the upper cut-off of the distribution and an index $q$ parameterizing the time evolution of the cut-off will also be included as fit parameters \citep{ 2008MNRAS.388..144R}. In the forward-reverse shock model, the spectral parameters of the reverse shock and the shock crossing time $t_x$ are also included as free parameters. In the two-component model we have the deceleration time $t_{\rm{dec}}$ of the wide jet as an additional free parameter. A band type smoothening is used in spectral (between multiple powerlaws of the synchrotron spectrum) and in the temporal (for gradual transition of the spectral parameters across $t_j$, $t_{\rm{dec}}$ and $t_x$) domains \citep{2002ApJ...568..820G}. We derive the physical parameters from the best fit spectral parameters. For the wind termination shock model,  we use the afterglow code of \cite{2006ApJ...647.1238J} to calculate the lightcurves. It takes as input the physical parameters $E_{\rm iso}$, $A_*$, $\Delta n$, $R_{sw}$, $p$, $\epsilon_e$, and $\epsilon_b$, where $\Delta n$ is the change in density at the wind termination shock and $R_{sw}$ 
is the radius of the termination shock. In addition to these parameters, we also fit for dust extinction $A_V$ in the host frame. The multi-wavelength fit therefore directly provides the best fit physical parameters.  Note that, in this model the definition of $\epsilon_e$ has been changed from the one used in \citet{2006ApJ...647.1238J} to that of \citet{2001ApJ...554..667P} to allow for values of $p<2$. 

\subsection{Forward-reverse shocks}
\cite{2005ApJ...633.1027S} modelled the {\it R}-band light curve alone using a forward-reverse shock model. 
In this model, for $T < $0.02~days the RS dominates the emission. 
The rise at $T\sim$0.02~days coincides with the passage 
of the synchrotron frequency of the forward shock through {\it R}-band, after which the FS dominates the light curve.
However,  in the {\it Swift} {\it V}-band,  with a good sampling at $T <$0.02~days, the light curve decay is too shallow to be explained by standard reverse shock emission. Nevertheless we first attempted an forward-reverse shock model where the optical light curve at $T <$0.02~days is not entirely dominated by the RS, but also by the rising FS. This allowed us to have the steep decay of the RS, but still reproduce the early optical decay. However, this placed constraints on the temporal profile of the modelled FS emission at the early period. To reproduce the observed optical and X-ray light curves, we had to have a wind driven density profile for the ambient medium, and a hard ($p<2$) electron distribution \citep{2001BASI...29..107B, 2001BASI...29...91S, 2005BASI...33..487M, 2008MNRAS.388..144R} undergoing fast cooling during this time period. We do not have to include dust extinction in the host frame to achieve this fit. Along with further constraints on the FS emission from the later part of the optical and X-ray light curves, and the {\it Spitzer} spectrum \citep{2008ApJ...681.1116H} at $2.3$~days, we narrowed down the parameter space.  In Figure \ref{figure4}, we present the best fit model (reduced chi squared,  $\chi^2_{\rm{DOF}} = 3.9$ for DOF of 223) 
along with the optical, X-ray and radio data.  We obtained a good fit in the spectral parameter space 
but the physical parameters inferred from these spectral fits turned out to be unrealistic (Table \ref{table3}) and based on that we had to rule out this model.  

\subsection{Two-component outflow}
We attempt a two-component model, where different outflows dominate the early and late-time light curves. The most popular picture is of two co-aligned components with different opening angles  \citep{2003Natur.426..154B, 2008Natur.455..183R}. 
For GRB 050525A, the first (narrow) jet dominates the emission at $T < 0.02$~days while the second (wide) jet gives rise to the $R$-band re-brightening at $T\sim  0.02$~days as it decelerates around this epoch, and dominates thereafter.
The late-time optical and X-ray light curves are seen to steepen gradually. We assume this to be due to the lateral expansion of the wide component and our best fit jet-break is at $\sim 5$~days. Due to the lack of dense sampling around this time, there are relatively larger uncertainties on this estimate. After $T > 0.1$ days the emission from the narrow component is required to decay faster so that the model does not over-predict the flux at later epochs. Even though various reasons could cause the quenching, the most natural assumption is that the narrow component is undergoing lateral expansion around that time. The narrow jet contribution is significant in X-rays for sufficiently long time and hence relatively good constraints on its $t_j$ can be obtained from the X-ray light curve alone. A constant density ambient medium profile produced the best fit. The model required a Milky Way type additional extinction in the host galaxy frame.

In Figure \ref{figure5}, we present the best fit model 
along with the optical, X-ray and radio data. We obtain a somewhat high $\chi^2_{\rm{DOF}}$ of $4.5$ (for DOF of 223). Such large values are due to the small scale fluctuations in the light curves, especially the scatter around the re-brightening epoch in $R$-band and the early flare-like feature in X-ray. Our best fit spectral and physical parameters are given in Table \ref{table4}. Our fit parameters (especially the value of $p$, the host extinction and $t_j$ for the second jet) differ from the values reported in previous studies \citep{2006ApJ...637..901B,2006ApJ...642L.103D,2008ApJ...681.1116H}, because our numerical code uses a more sophisticated description than the multiple powerlaw fitting employed in these papers. Since the narrow component is relevant only for a short period of time, its spectral parameters are not well constrained.  We are only able to obtain lower limits to the self absorption frequency ($\nu_a$) and cooling frequency ($\nu_c$). The inferred physical parameters are well within the range observed for other afterglows 
(\citealt{2001ApJ...560L..49P, 2008MNRAS.388..144R, 2010ApJ...711..641C}). The ambient medium density is towards the higher end of what is normally estimated in afterglow modelling, leading to a wider inferred collimation angle for the given value of $t_j$. The estimated value of kinetic energy in both the jets are similar. The initial lorentz factor ($\Gamma_0$) of the wide jet is $\sim 10$ and that of the narrow jet is $ > 60$, suggesting the more relativistic narrow jet to be responsible for the prompt gamma-ray emission. The energy in the narrow jet is $\sim 370$ times that in the radiation observed by BAT ($2.3 \times 10^{52}$~erg, \citep{2006ApJ...637..901B}). This could indicate that only around $0.3 \%$ of total energy is converted to radiation during the prompt emission phase, which is compatible with the low efficiencies expected for internal shocks \citep{1998MNRAS.296..275D}.  

Since for both jets $\epsilon_e \gg \epsilon_B$, we estimated the expected inverse compton emission and found it to be significantly lower than the synchrotron component.

The model agrees reasonably well with the VLA radio data in multiple frequencies, however the radio light curves are more or less flat for the entire duration of VLA observation. The $8.4$~GHz and $4.8$~GHz fluxes remain at the same level within error bars from  $0.3$ to $30$ days. 
This could be caused either by contamination from the associated supernova or by contribution from a nearby unresolved variable source (like an AGN).
We note that a similar flattening has been observed in other radio afterglows \citep{2004ApJ...600..828F}. 
Nevertheless, we have provided the best possible fits to the available radio data within our model. The steepening in the light curves corresponds to the epoch when the fireball becomes optically thin to synchrotron radiation.

\subsection{Wind driven bubble and termination shock}
The two-component model agrees well with the overall behaviour of the multi-wavelength light curves. However,  for the early X-ray light curve ($0.001--0.012$~days),  only the average behaviour is explained by this model.  The X-ray light curve during this period displays mild flares. In an attempt to explain this finer feature, we opted for another model where the ambient density profile changes shape, such as in a wind bubble. The immediate neighbourhood of the burst has a stratified density profile up to the termination shock $R_{\rm{sw}}$, where the afterglow shock encounters a density jump as well as a change in the density profile. When the forward shock enters the density enhancement it rapidly slows down and the co-moving density of the shocked matter is increased. There is therefore an increased number of radiating electrons but their average energy is lower.  This causes the afterglow temporal decay to be slower for $\nu  < \nu_c$ but faster for $\nu >\nu_c$ because $\nu_c$ rapidly decreases when the shock enters the density jump \citep{JohPhDThesis}. The afterglow then settles into a powerlaw decay   
in the constant density medium. If $\nu_c$ remains between the optical and X-ray observing bands in this afterglow, this model can reproduce the optical bump observed in the light curve.

Our best fit model is shown in Figure \ref{figure6} where the wind termination shock is placed at $\sim
0.09$~pc (corresponding to the observer frame time $T \sim 900$~s), and has a density increase of a factor of $\sim 12$.  The set of parameters used for the model are given in Table~\ref{table5}.  
The model parameters differ from those determined from the two-component model given in 
Table \ref{table4} because of the different modeling approach.  The parameters are more like the narrow component in the two-component model,  in agreement with the fact that it better fits the early afterglow data than the late data. The typical distance to a WR stars termination shock ranges from $\sim 0.1$ pc to a few hundred pc depending on the mass loss rate of the star and the interstellar medium surrounding it
\citep{2006MNRAS.367..186E, 2006A&A...460..105V, 2007RSPTA.365.1255E}.
The radius of the wind termination shock is much smaller than that expected for a typical 
Wolf-Rayet star, possibly indicating that the afterglow erupted from the 
rotational axis and the star was in a dense interstellar environment \citep{2007MNRAS.377L..29E}.  

While the model adequately describes the optical and infrared light curves and early 
X-ray light curve, it over-predicts the late-time X-ray light curve and 
under-predicts the late-time radio observations. A possible explanation for this 
difference is a change in the shock physics causing the cooling break frequency to 
decrease further, for example a small increase in the magnetic field strength of the 
shock. Further density fluctuations might also explain the differences as 
numerical modelling have shown the wind of massive stars to be more structured
than we assume in our model \citep{2007MNRAS.377L..29E}. We obtain a $\chi^2_{\rm{DOF}}$ of $\sim 3.6$ (for a DOF of 223). For this model too, the scatter in {\it R}-band around the rebrightening epoch contributes to the somewhat large value of $\chi^2$.

\subsection{Supernova contribution} 
\label{sn}
\begin{figure}
\begin{center}
\includegraphics[scale=0.9]{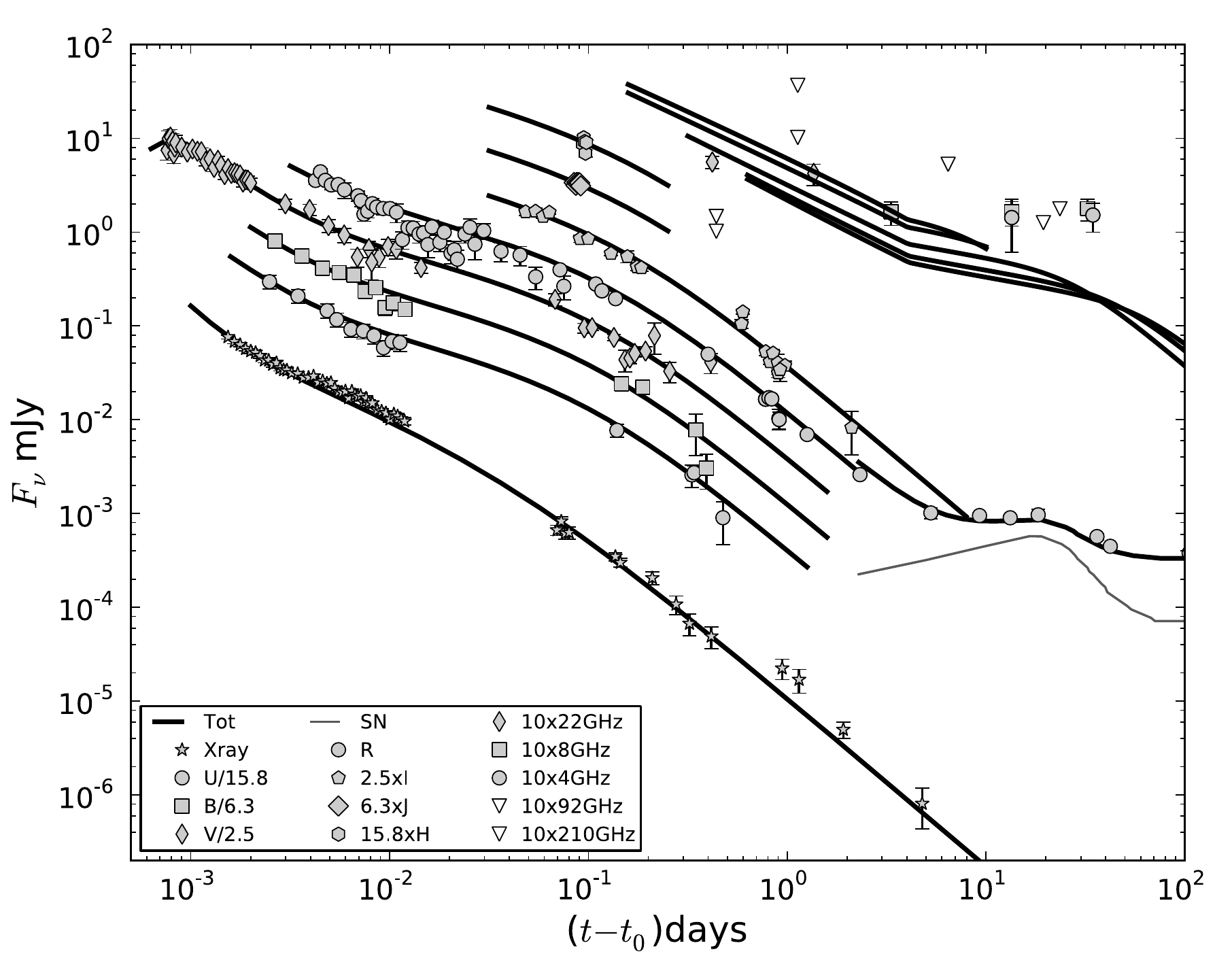}
\end{center}
\caption{Optical {\it UBVRIJH},  {\it Swift} X-ray,  radio and millimeter light curves of the afterglow of GRB~050525A along with the forward-reverse shock model.  The supernova component is shown as a thin gray line.  The light curves are shifted arbitrarily for clarity.  XRT data is recovered from \citet{2006ApJ...637..901B} and the VLA radio data from the NRAO repository.}
\label{figure4}
\end{figure}

\begin{figure}
\begin{center}
\includegraphics[scale=0.9]{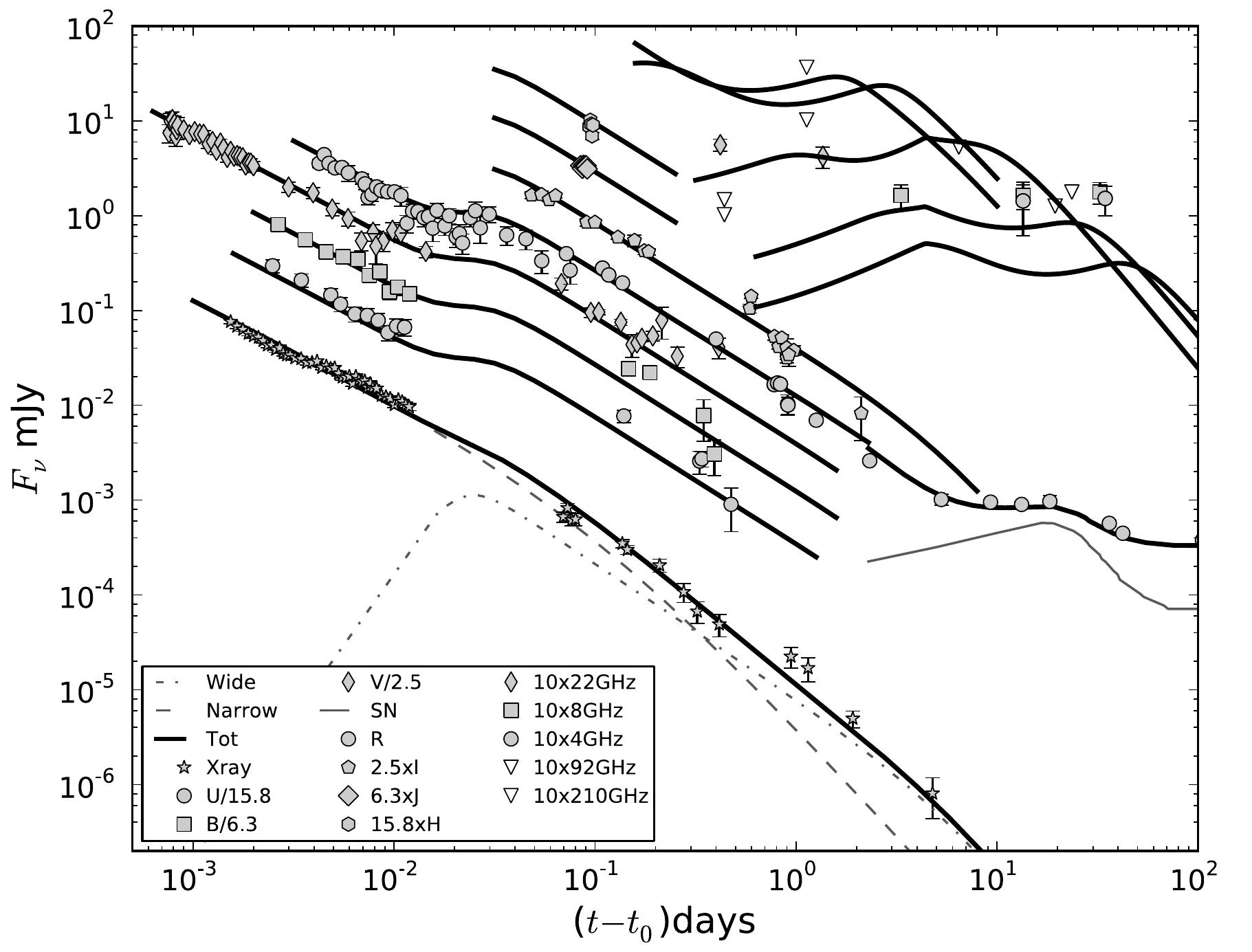}
\end{center}
\caption{Same as in figure \ref{figure4} but for the best fit two-component outflow model.  For the X-ray light curve, contribution from the narrow and wide components are shown as dashed and dash-dotted curves separately.}
\label{figure5}
\end{figure}

\begin{figure}
\begin{center}
\includegraphics[scale=0.9]{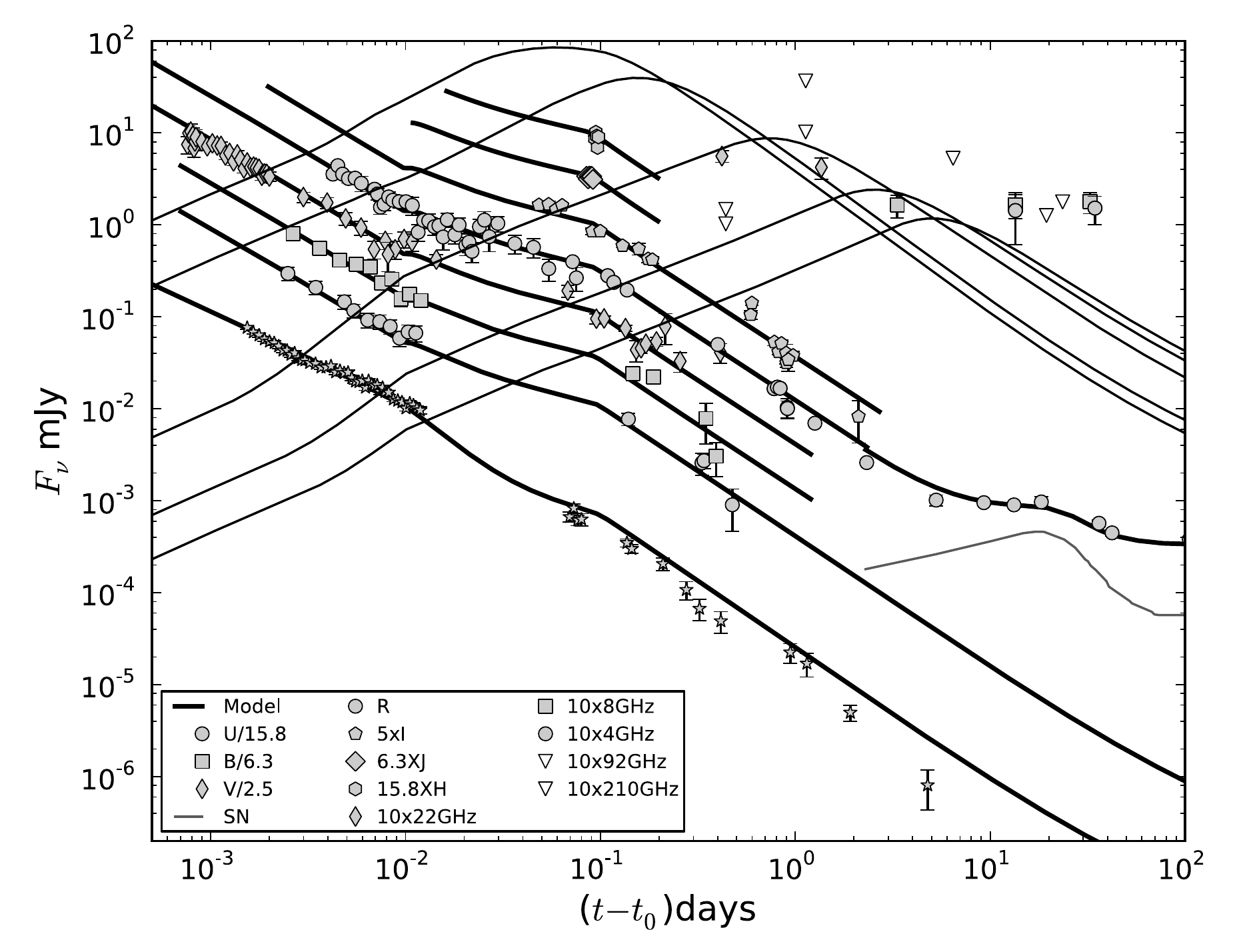}
\end{center}
\caption{Same as in figure \ref{figure5} but for the best fit wind termination shock model.}
\label{figure6}
\end{figure}

Optical light curves of several low-redshift GRBs show a late-time red bump \citep{2004ApJ...609..952Z}, often attributed to the emission from a supernova believed to be associated with the progenitor of the GRB \citep{2003ApJ...591..288H}. Around $T \sim 5$~days, a flattening of the $R$-band light curve was observed for GRB 050525A \citep{2006ApJ...642L.103D} with a temporal profile similar to the `SN bump' seen in other afterglows. The late-time ($\sim 36$~days) VLT spectrum observed by \citeauthor{2006ApJ...642L.103D} also showed similarities to the spectra of other GRB associated SNe. We compared the late-time $R$-band light curve with the prototypical SN98bw \citep{1998Natur.395..670G}, the first known example of a supernova associated with a GRB. We used the template optical light curves of SN98bw \citep{1998Natur.395..670G} k-corrected for the redshift of GRB 050525A. Best agreement with the observed $R$-band data for $T > 5$~days were obtained after shifting the SN98bw template by $-0.9$ days and applying a temporal stretch factor of $0.8$. We did not have to adjust the magnitude of the shifted SN98bw, implying the associated SN to be of similar brightness as SN98bw. The host galaxy $R$-band magnitude was fixed at $25.1$ \citep{2006ApJ...642L.103D}. The SN component is shown as thin gray lines in Figures \ref{figure4},  \ref{figure5} and \ref{figure6}.

%
%
%

\section{Summary}
\label{conclusion}
We have presented the optical afterglow of GRB~050525A in {\it VRIJH} photometric bands. Our data fill some gaps in the optical multi-wavelength light curves beyond $0.01$~days and provide a better constraint on the optical decay index. 
Our {\it IR} observations, though confined to a narrow time bin, provide additional spectral constraints. The millimeter upper limits contributes to a better picture of the low frequency behaviour of the afterglow.

We have undertaken a comprehensive multi-wavelength modelling of the afterglow,  and tested various models against the data. The afterglow behaviour is too complex for a simple blast-wave model. We find that including emission from a possible reverse shock component is not sufficient to explain the afterglow evolution. Either the outflow is structured as a two-component jet or the ambient medium has a complex structure involving a variation in the density profile. Our two-component jet model is able to reproduce the overall behaviour of the afterglow, except the finer fluctuation in the early X-ray light curve. The wind termination shock model succeeds in explaining the early phase including the short time-scale features but deviates from the late-time data. The radio light curves are moderately well explained by both models. Afterglow modelling is necessary to unravel the nature of the outflow and the structure of the medium around the burst. The complexity of the observed light curves demands that the most realistic models are used.  Dense sampling and long monitoring campaigns are also required in conjunction with afterglow modelling.   

\section*{Acknowledgments}
We are thankful to an anonymous referee for useful comments and suggestions which has improved the presentation of the paper. We thank and acknowledge the observing support at different observatories in collecting the data. This research has made use of data obtained through the High Energy Astrophysics Science Archive. We thank the VLA staff for easy access to the archival data base. The National Radio Astronomy Observatory is a facility of the National Science Foundation operated under co-operative agreement by Associated Universities Inc. Research Center Online Service, provided by the NASA/Goddard Space Flight Center. This work contains observations carried out with the IRAM Plateau de Bure Interferometer. IRAM is supported by INSU/CNRS (France), MPG (Germany) and IGN (Spain). We thank the support of the Spanish MICINN projects AYA2009-14000-C03-01 and AYA2008-03467/ESP. The DARK Cosmology Center is funded by the DNRF. \\
%
%
%

%
%
%
\begin{table*}
\caption{Log of Millimeter wave observations}
\centering
\smallskip
\begin{tabular}{c c r c c c}
\hline \hline
Date-obs (UT)            & Time since burst ($T$)$^\dagger$  & Frequency &  3-$\sigma$ upper & Synthesised Beam                 & Position Angle\\
(d)                      &      (d)                          & (GHz)     &  limit (mJy)      &                                  &  (deg)\\
\hline                                                                     
2005-05-25.9188--26.3479 & 0.9168--1.3459                    & 92.682    & $<$1.02              & 7$\farcs$05 $\times$ 5$\farcs$58 & 74 \\
	               	 & 		                     & 214.712   & $<$3.66              & 3$\farcs$10 $\times$ 2$\farcs$36 & 78 \\
\hline                                    
\end{tabular}
\newline
$^\dagger$ $T=t-t_0$ where $t_0=$25-05-25.002 d 
\label{table1}      
\end{table*}

\begin{table*}
\caption{Log of optical and NIR observations}
\centering
\smallskip
\begin{tabular}{l l c c c l}
\hline \hline
Date-obs (UT)      & Time since burst ($T$)$^\dagger$ & Magnitude    & Exp time &  Filter & Telescope\\
 (d)               &    (d)                     & (mag)        &  (s)     &         &          \\
\hline
2005-05-25.0101  & 0.0081 & 16.510 $\pm$ 0.390  &   10$\times$32   &   $V$ &   0.2m-Bootes1-B\\
2005-05-25.0682  & 0.0680 & 17.500 $\pm$ 0.165  &   300   &   $V$ &   1.2m-Mercator\\
2005-05-25.0955  & 0.0953 & 18.260 $\pm$ 0.134  &   300   &   $V$ &   1.2m-Mercator\\
2005-05-25.1046  & 0.1044 & 18.253 $\pm$ 0.072  &   300   &   $V$ &   1.2m-Mercator\\
2005-05-25.1347  & 0.1344 & 18.517 $\pm$ 0.082  &   300   &   $V$ &   1.2m-Mercator\\
2005-05-25.1528  & 0.1526 & 19.107 $\pm$ 0.293  &   300   &   $V$ &   1.2m-Mercator\\
2005-05-25.1619  & 0.1616 & 19.063 $\pm$ 0.142  &   300   &   $V$ &   1.2m-Mercator\\
2005-05-25.1713  & 0.1711 & 18.941 $\pm$ 0.090  &   300   &   $V$ &   1.2m-Mercator\\
2005-05-25.1940  & 0.1937 & 18.878 $\pm$ 0.128  &   300   &   $V$ &   1.2m-Mercator\\
2005-07-01.0386  & 37.0366& $>$22.5             &  1000$\times$5    &   V &  2.5m-INT\\
&&&&&\\                                                             
2005-05-25.1092  & 0.1089 & 17.790 $\pm$ 0.054  &   300   &   $R$ &   1.2m-Mercator\\
2005-05-25.1137  & 0.1169 & 17.969 $\pm$ 0.043  &   900   &   $R$ &   1.2m-Mercator\\
2005-05-25.7800  & 0.7822 & 20.860 $\pm$ 0.079  &   720   &   $R$ &   2.0m-HCT\\
2005-05-25.8108  & 0.8123 & 20.819 $\pm$ 0.070  &   600   &   $R$ &   2.0m-HCT\\
2005-05-25.8376  & 0.8426 & 20.850 $\pm$ 0.079  &  1200   &   $R$ &   2.0m-HCT\\
2005-05-25.9111  & 0.9143 & 21.370 $\pm$ 0.270  &   900   &   $R$ &   1.0m-ST\\
2005-05-25.9130  & 0.9144 & 21.399 $\pm$ 0.230  &   600   &   $R$ &   2.0m-HCT\\
2011-05-26.1508--30.0955  & 2131.1488--2135.0935  &  $<$24.3               & 18$\times$900   &  $r$      & 2.2m-CAHA\\
&&&&&\\                                                             
2005-05-25.0485  & 0.0483 & 16.533 $\pm$ 0.093  &   300   &   $I$ &   1.2m-Mercator\\
2005-05-25.0544  & 0.0541 & 16.517 $\pm$ 0.097  &   300   &   $I$ &   1.2m-Mercator\\
2005-05-25.0594  & 0.0591 & 16.656 $\pm$ 0.043  &   300   &   $I$ &   1.2m-Mercator\\
2005-05-25.0638  & 0.0636 & 16.547 $\pm$ 0.056  &   300   &   $I$ &   1.2m-Mercator\\
2005-05-25.0910  & 0.0907 & 17.249 $\pm$ 0.098  &   300   &   $I$ &   1.2m-Mercator\\
2005-05-25.1001  & 0.0998 & 17.247 $\pm$ 0.043  &   300   &   $I$ &   1.2m-Mercator\\
2005-05-25.1301  & 0.1299 & 17.643 $\pm$ 0.054  &   300   &   $I$ &   1.2m-Mercator\\
2005-05-25.1574  & 0.1571 & 17.726 $\pm$ 0.159  &   300   &   $I$ &   1.2m-Mercator\\
2005-05-25.1758  & 0.1756 & 18.002 $\pm$ 0.098  &   300   &   $I$ &   1.2m-Mercator\\
2005-05-25.1849  & 0.1847 & 18.024 $\pm$ 0.090  &   300   &   $I$ &   1.2m-Mercator\\
2005-05-25.5918  & 0.5915 & 19.520 $\pm$ 0.119  &   300   &   $I$ &   1.0m-ST\\
2005-05-25.6002  & 0.5999 & 19.200 $\pm$ 0.059  &   300   &   $I$ &   1.0m-ST\\
2005-05-25.7826  & 0.7842 & 20.260 $\pm$ 0.100  &   630   &   $I$ &   2.0m-HCT\\
2005-05-25.8216  & 0.8241 & 20.530 $\pm$ 0.070  &   780   &   $I$ &   2.0m-HCT\\
2005-05-25.8514  & 0.8563 & 20.299 $\pm$ 0.100  &  1200   &   $I$ &   2.0m-HCT\\
2005-05-25.8959  & 0.9008 & 20.829 $\pm$ 0.119  &  1200   &   $I$ &   2.0m-HCT\\
2005-05-25.8995  & 0.9028 & 20.559 $\pm$ 0.270  &   900   &   $I$ &   1.0m-ST\\
2005-05-25.9659  & 0.9744 & 20.627 $\pm$ 0.129  &  1800   &   $I$ &   1.5m-OSN\\
2005-05-25.9223  & 0.9256 & 20.739 $\pm$ 0.280  &   900   &   $I$ &   1.0m-ST\\
2005-05-27.1116  & 2.1096 & 22.390 $\pm$ 0.540  &  200$\times$12  &   $I$ &   1.5m-OSN\\
2005-06-30.9721  & 36.048 & $>$23               & 450$\times$30   &   $I$ &   2.2m-CAHA\\
2011-05-28.9855--30.0955  & 2133.9845--2135.0935  & 25.1$\pm$0.4           & 17$\times$900   &  $i$      & 2.2m-CAHA\\
&&&&&\\                                                             
2011-05-26.1508--30.0955  & 2131.1488--2135.0935  &  $<$24.1               & 18$\times$900   &  $g$      & 2.2m-CAHA\\
&&&&&\\                                                             
2005-05-25.0953  & 0.0936 & 15.779 $\pm$ 0.068  &  60.240 &   $H$ &   3.5m-CAHA\\
2005-05-25.0963  & 0.0946 & 15.611 $\pm$ 0.072  &  60.240 &   $H$ &   3.5m-CAHA\\ 
2005-05-25.0973  & 0.0956 & 15.713 $\pm$ 0.064  &  60.240 &   $H$ &   3.5m-CAHA\\ 
2005-05-25.0983  & 0.0966 & 16.024 $\pm$ 0.093  &  60.240 &   $H$ &   3.5m-CAHA\\ 
2005-05-25.0992  & 0.0976 & 15.737 $\pm$ 0.067  &  60.240 &   $H$ &   3.5m-CAHA\\ 
&&&&&\\                                                             
2005-05-25.0866  & 0.0847 & 16.263 $\pm$ 0.039  &  21.079 &   $J$ &   3.5m-CAHA\\
2005-05-25.0881  & 0.0865 & 16.337 $\pm$ 0.034  &  60.240 &   $J$ &   3.5m-CAHA\\
2005-05-25.0891  & 0.0875 & 16.257 $\pm$ 0.035  &  60.240 &   $J$ &   3.5m-CAHA\\
2005-05-25.0901  & 0.0884 & 16.294 $\pm$ 0.034  &  60.240 &   $J$ &   3.5m-CAHA\\
2005-05-25.0911  & 0.0894 & 16.263 $\pm$ 0.034  &  60.240 &   $J$ &   3.5m-CAHA\\
2005-05-25.0920  & 0.0904 & 16.271 $\pm$ 0.034  &  60.240 &   $J$ &   3.5m-CAHA\\
2005-05-25.0931  & 0.0914 & 16.353 $\pm$ 0.035  &  60.240 &   $J$ &   3.5m-CAHA\\
2011-07-18.0320--18.1150  & 2245.0301--2245.1131  & 23.58$\pm$0.225$^{\dagger\dagger}$ & 83$\times$60    &  $J$      & 3.5m CAHA\\
2011-07-19.0502--19.1077  & 2246.0483--2246.1059  &    --                              & 57$\times$60    &  $J$      & 3.5m CAHA\\
\hline                                   
\end{tabular}
\newline
$^\dagger$ $T=t-t_0$ where $t_0=$25-05-25.002 d 
\newline
$^{\dagger\dagger}${The magnitude is the result of the combination of the $J$-band data taken on 18 and 19 July 2011.}
\label{table2}      
\end{table*}

\begin{table*}
\caption{Best fit ($\chi^2_{\rm{DOF}}\sim 3.9$) spectral and physical parameters of the forward-reverse shock model. No host extinction need to be used. However, the values of the derived physical parameters are unrealistic and hence we rule out this model.}
{\scriptsize
\begin{tabular}{c|c|c|c|c|c|c|c|}
\hline \hline
&$\nu_a$ (Hz) & $\nu_m$ (Hz) & $\nu_c$ (Hz) &  $f_m$ (mJy) & $p$ & $q$ & $t_j$~(day)  \\ 
{\rotatebox{90}{shock}}&$\sim 10^{10}$  & $8.0^{+4.5}_{-5.5} \times 10^{15}$ & $7.08^{+5.5}_{-1.5} \times 10^{11}$  & $15.82^{+4.1}_{-1.72}$ & $1.82^{+0.08}_{-0.18}$ & $1.2 \pm 0.2$ & $0.04^{+0.01}_{-0.02}$     \\ 
&&&&&&\\
\cline{2-8}
{\rotatebox{90}{Fwd.}}&$E_{\rm{iso,52}}$ (erg) & $A_*$ & $\epsilon_e$ & $\epsilon_B$ & $\theta_j \, (^{\circ})$ & $E_{\rm{tot,52}}$~(erg) & \\
&&&&&&\\
&$11.75^{+0.21}_{-0.19} $ & $2.63^{+2.9}_{-0.71} \times 10^{-6}$ & $1.6^{+0.9}_{-0.7} \times 10^{-3} \Lb\frac{\nu_i}{3 \times 10^{19} Hz}\Rb^{0.12}$ & $(2.5 \pm 1.8) \times 10^{7}$ & $0.11^{+0.03}_{-0.02}$ & $2.^{1.2}_{-0.7} \times 10^{-5}$ & \\ 
&&&&&&\\
\hline
&$\nu_a$ (Hz) & $\nu_m$ (Hz) & $\nu_c$ (Hz) &  $f_m$ (mJy) & $p$ & $t_x$~(day) &  \\
{\rotatebox{90}{shock}}& --- & $2.^{+2.}_{-0.74}\times 10^{11}$ & $2.95^{+2.94}_{-1.47} \times 10^{15}$  & $10.^{+1.2}_{-2.1}$ & $2.08 \pm 0.07$ & $7.9^{+1.}_{-0.86} \times 10^{-4}$   &   \\ 
&&&&&&\\
\cline{2-8}
{\rotatebox{90}{Rev.}}& & & $\epsilon_e$ & $\epsilon_B$ & $\Gamma_0$ & &  \\
&&&&&&\\
&&& $\sim 1.1$ & $(1. \pm 0.2)\times 10^{-3}$ &$1073^{+116}_{-204}$& & \\
\hline
\end{tabular}
}
\label{table3}
\end{table*}

\begin{table*}
\caption{Best fit spectral and physical parameters of the two-component outflow model. The best fit $A_v$(host) is $0.16 \pm 0.06$ mag (Milky Way type). The spectral parameters are quoted at the jet-break time of the respective components.  Variabilities in the early optical and X-ray light curves result in a large value of $\chi^2_{\rm{DOF}}\sim 4.5$.}
{\scriptsize
\begin{tabular}{|c|c|c|c|c|c|c|c|}
\hline \hline
&$\nu_a$ (Hz) & $\nu_m$ (Hz) & $\nu_c$ (Hz) & $f_m$ (mJy) & $p$& $t_j$ (day) &$t_{\rm{dec}}$ (day) \\ 
{\rotatebox{90}{component}}&$ (4.6^{+0.1}_{-0.4}) \times 10^{10}$ & $(3.5^{+1.1}_{-0.8}) \times 10^{10}$ & $(4.0 \pm 3.0) \times 10^{16}$ & $2.23^{+2.6}_{-1.7}$ & $2.6 \pm 0.08$ & $5.0 \pm 3.0$ & $0.025$ \\
&&&&&&&\\
\cline{2-8}
{\rotatebox{90}{wide}}&$E_{\rm{iso,52}}$ (erg)&$n_0$ (atom/cm$^{3}$)&$\epsilon_e$&$\epsilon_B$& $\theta_j$ ($^{\circ}$)&$E_{\rm{tot,52}}$ (erg) & $\Gamma_0$ \\
&&&&&&&\\
&$9.35^{+10.3}_{-9.0}$ & $354.8^{267.7}_{-154.6}$ & $0.126 \pm 0.02$ & $(5.1 \pm 3.4) \times 10^{-5}$ & $(24.8^{+4.7}_{-0.7})$ & $0.43 \pm 0.25$ & $7.1 \pm 2.1$ \\ 
&&&&&&&\\
\hline
&$\nu_a$ (Hz) & $\nu_m$ (Hz) & $\nu_c$ (Hz) & $f_m$ (mJy) & $p$& $t_j$ (day) & $t_{\rm{dec}}$ (day) \\ 
{\rotatebox{90}{component}}&$> 6.3 \times 10^{11}$ & $(1.0^{+1.1}_{-0.1}) \times 10^{11}$ & $ > 2.6 \times 10^{19}$ & $23.72^{+1.7}_{-2.6}$ & $2.37 \pm 0.02$ & $0.1^{+0.02}_{-0.01}$ &  $> 7.5 \times 10^{-4}$ \\
&&&&&&&\\
\cline{2-8}
&&&&&&&\\
&$E_{\rm{iso,52}}$ (erg)&$n_0$ (atom/cc)&$\epsilon_e$&$\epsilon_B$& $\theta_j$ ($^{\circ}$) &$E_{\rm{tot,52}}$ (erg)&$\Gamma_0$ \\
{\rotatebox{90}{narrow}}& $852.56 \Lb\frac{\nu_c}{3 \times 10^{19}}\Rb^{1/4}$ & $354.8$ & $0.014\Lb\frac{\nu_c}{3 \times 10^{19}}\Rb^{1/4} $ & $6.1 \times 10^{-6}\Lb\frac{\nu_c}{3 \times 10^{19}}\Rb^{-5/4}$ & $< 3.26$ & $0.69\Lb\frac{\nu_c}{3 \times 10^{19}}\Rb^{1/4}$ & $> 60$ \\ 
&&&&&&&\\
\hline
\end{tabular}
}
\label{table4}
\end{table*}

\begin{table}
\caption{Parameters of the wind termination shock model.  For explanation of parameters see \citep{2006ApJ...647.1238J}.  Note that the definition of $\epsilon_e$ has been changed to allow for values of $p < 2$ as in \citep{2001ApJ...554..667P}. Values without uncertainties were fixed in the model and do not affect the results significantly. Dust extinction in the host frame is $A_v = 0.15 \pm 0.05$ (SMC type). The best fit model yields a $\chi^2_{\rm{DOF}}\sim 3.6$.}
\centering
\smallskip
\begin{tabular}{r c}
\hline \hline
Parameter & Value \\
\hline
$E_{\rm{tot,52}}$ (erg)   & $0.1_{-0.08}^{+0.9}$ \\ 
$\theta_j$ ($^{\circ}$) & $< 1.1$ \\    
$A_*$                     & $0.5_{-0.45}^{+10}$ \\ 
$\Delta n ^{\dag}$        & $> 12$ \\ 
$R_{sw} ^{\ddag}$ (pc)    & $0.09 \pm 0.06$ \\ 
$p$                       & $1.85 \pm 0.15$ \\ 
$\epsilon_e/10^{-5}$      & $1_{-0.9}^{+5}$ \\
$\epsilon_B/10^{-4}$      & $3_{-2.8}^{+9}$ \\ 
& \\
\hline
\end{tabular}
\newline
$^\dag$ Fractional change in density at the termination shock \\
$^\ddag$ Radius of the termination shock
\label{table5}      
\end{table}


%
\end{document}